\documentclass{article}

\usepackage{arxiv}

\usepackage[utf8]{inputenc} 
\usepackage[T1]{fontenc}    
\usepackage{hyperref}       
\usepackage{url}            
\usepackage{booktabs}       
\usepackage{amsfonts}       
\usepackage{nicefrac}       
\usepackage{microtype}      
\usepackage{lipsum}
\usepackage{graphicx}
\graphicspath{ {./images/} }

\title{Tunable THz  Switch-Filter Based on Magneto-Plasmonic Graphene Nanodisk}

\author{
Victor Dmitriev\thanks{Use footnote for providing further
    information about author (webpage, alternative
    address)---\emph{not} for acknowledging funding agencies.} \\
 Department of Electrical Engineering\\
  Federal University of Para\\
  66075-900, Belem, Para, Brazil. \\
  \texttt{victor@ufpa.br} \\
   \And
 Geraldo Melo\\
  Federal Rural University of Amazonia\\
68700-030, Campus Capanema, Para, Brazil.\\
    \texttt{geraldo.melo@ufra.edu.br} \\
  \And
 Wagner Castro\\
  Institute Cyberspace\\
  Federal Rural University of Amazonia\\
 66.077-830, Belem, Para, Brazil.\\
    \texttt{wagner.ormanes@ufra.edu.br} \\
}

\begin{document}
\maketitle
\begin{abstract}
We propose and analyze a multifunctional THz device  which can operate as a tunable switch and a filter.   The device consists of a  circular graphene nanodisk coupled to  two nanoribbons  oriented at  $90^\circ$ to each other.  The graphene elements are placed on a dielectric substrate. The nanodisk  is magnetized by a DC  magnetic field normal to its plane. The physical principle of the device is based on the propagation of surface plasmon-polariton waves  in the graphene nanoribbons and excitation of  dipole modes in the nanodisk. Numerical simulations show that  0.61T DC magnetic field provides transmission (regime ON) at the frequency 5.33 THz with   the bandwidth 12.7$\%$ and filtering with the Q-factor equals to 7.8. At the central frequency, the insertion loss is around -2 dB and the  reflection coefficient is -43 dB. The regime OFF can be achieved by means of switching DC magnetic field to zero value or by switching chemical potential of the nanodisk to zero with the ON/OFF ratio better than  20 dB. A small central frequency tuning by chemical potential  is possible with a fixed DC magnetic field.
\end{abstract}


\section{Introduction}
After discovery of graphene in 2004 \cite{graphene},  many new photonic and electronic components  and, in particular, multifunctional devices based  on  graphene have been suggested  \cite{R1,R2,R3}. Switch is one of the key elements of modern digital technology used  in majority of the circuits. Depending on its state (ON/OFF), the switch can turn on or block some function of the circuit. 

Today, there exist many types of switches based on different physical effects. The physical principles used in switches depend on the  frequency region and the power of electromagnetic waves. For example, in high power microwave systems, mechanical switches \cite{Switches1} are used. For low power systems, especially in microstrip technology, semiconductor switches are common circuit elements \cite{Switches2}. Electro-optical, acusto-optical, magneto-optical and nonlinear effects are used for switching and amplitude modulation \cite {S3,S4,S5} in microwaves and  optics. In particular, electro-optical Mach-Zehnder interferometer is a typical elements in optical circuits \cite {Mach-Zehnder}. In photonic crystal technology, several types of switches were also suggested \cite {photonic-crystal}.

In THz region, several solutions for switch technology are already being suggested, among them  structures based on metals and dielectrics, graphene and photonic crystals \cite {Feng, Zhang, Luo, Ali, Sun, Guo}. One can also find in the literature examples of switches based on magnetized graphene \cite{y, MP}. Under the action of external DC magnetic field   graphene acquires new properties which can be explored for control and non-reciprocal devices. Different types of filters including graphene based tunable ones have been suggested in literature as well, see for example  \cite{Fi1,Fi2}.

In this paper, we suggest a new type of graphene-based component which can operate as a tunable switch  and a bandpass filter. The  switching mechanism can be provided by means of DC magnetic or electric field.  The theoretical results  are validated by full wave simulations using the commercial software Multiphysics version 5.2a \cite{comsolSite}. 
\section{Switch Description}

\begin{figure}[htbp]
\centering
\includegraphics[scale=0.8]{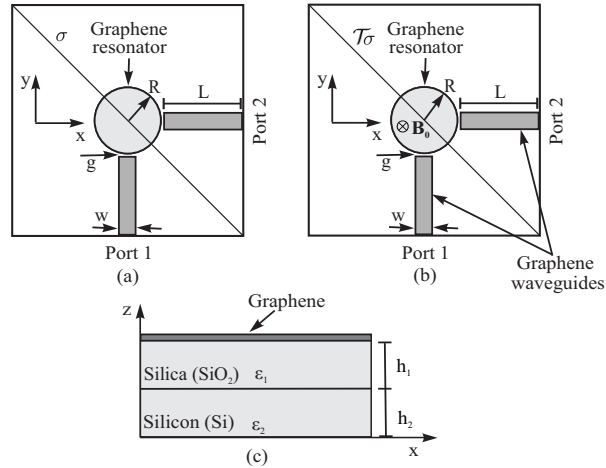}
\caption{Schematic representation of graphene device: (a) top view without magnetization, $\sigma$ is plane of symmetry;  (b) top view with magnetization, $\mathcal{T}\!\sigma$; (c) side view, ${B}_0$ is biasing DC magnetic field.}
\label{fig:Figdispositivo}
\end{figure}

Graphene nanoribbon with a waveguide-like propagating surface plasmon-polariton (SPP) mode  we shall call  further as graphene waveguide and  the nanodisk as graphene resonator. The main part of the suggested device is a circular graphene resonator with the radius $R$ which is coupled to two graphene waveguides with the widths $w$ and length $L$, see 
Fig. \ref{fig:Figdispositivo}.
The waveguides are oriented to each other at the angle $90^{\circ}$. There is small gap $g$ between the waveguides and the resonator. The graphene elements are placed on a substrate with a spacer defined by the thicknesses $h_1$ and $h_2$ and dielectric constants $\varepsilon_1$ and $\varepsilon_2$ respectively. 

From the point of view of magnetic biasing, the graphene disk can be in two sates, namely, without magnetization or it can be  magnetized by a DC magnetic field ${B}_0$ which is oriented  normally to the graphene plane. From the point of view of chemical potential, the graphene disk can also be in two state, i.e.  with biasing by electric field or without biasing. For small shifting of the central frequency of the device, the biasing electric field can be  controlled smoothly.

Symmetry of the device in nonmagnetized state is described by a geometrical plane of symmetry $\sigma$ (see Fig. \ref{fig:Figdispositivo}). With magnetization, it is described by the antiplane $\mathcal{T}\sigma$ where $\mathcal{T}$ is the time reversal operator  \cite{magnetic}.  The operator $\mathcal{T}$ appears due to the presence of the magnetic field ${B}_0$. Note that we use the so-called restricted time inversion operator, which preserves the passive or active nature of the medium.
\section{Optical Conductivity Tensor}
Surface conductivity $\sigma_s = (\omega,\mu_c({E}_0),\Gamma, \omega_{B})$  describes  interaction of graphene and electromagnetic radiation. It depends on   frequency $\omega$, chemical potential $\mu_{c}$ which is a function of biasing by electric field $E_{0}$, and also on phenomenological scattering rate, defined as $\Gamma = 1/\tau$, ($\tau$ relaxation time of graphene). In the magnetized state, $\sigma_s$ depends also on the cyclotron frequency $\omega_B ={eB_0v_{F}^{2}}/{\mu_c}$, where $e$ is the electron charge, $v_F$ is the Fermi velocity, $B_0$ is DC magnetic field. 

With applied field ${B}_0$, the moving charge carriers of graphene are controlled by Lorentz force. As a result, the 2D tensor of conductivity acquires anti-symmetric nondiagonal components \cite{sigma}:
\begin{equation}
[\sigma_s]=\left[\begin{array}{cc}
\sigma_{xx}  &\!\!\!\! -\sigma_{xy} \\
\sigma_{xy}  & \sigma_{xx}\\
\end{array} \right].
\end{equation}

It is known that the interband transitions in the THz frequency region can be neglected. In this case, one can use the classical Drude form of the tensor components \cite{inter}:
\begin{eqnarray}
\label{eq:sig1}
\sigma_{xx}=\frac{2D}{\pi} \frac{1/\tau-i\omega}{ \omega_{B}^{2}-( \omega + i / \tau ) ^2}, \\
\label{eq:sig2}
\sigma_{xy}=-\frac{2D}{\pi} \frac{\omega_c}{ \omega_{B}^{2}-( \omega+i / \tau )^2},
\end{eqnarray}
where $D=2\sigma_0\mu_c/\hbar$ is the Drude weight, $\sigma_0=e^2/(4\hbar)$ is the universal conductivity, $\mu_c$ is chemical potential of graphene, $\hbar$ is the reduced Planck's constant, $e$ is the electron charge, $\omega$ is  the angular frequency of the incident wave, $B_{0}$ is the DC magnetic field,  $i=\sqrt{-1}$ and $\tau=0.9$ ps is the relaxation time. 
The relaxation time is defined by $\tau=\mu\mu_c/(e v_F^2)$ where $\mu$ is electron mobility and $v_F\approx10^{6}$ m/s \cite{tau} is Fermi velocity. 

\begin{figure}[htbp]
\centering
\includegraphics[scale=0.9]{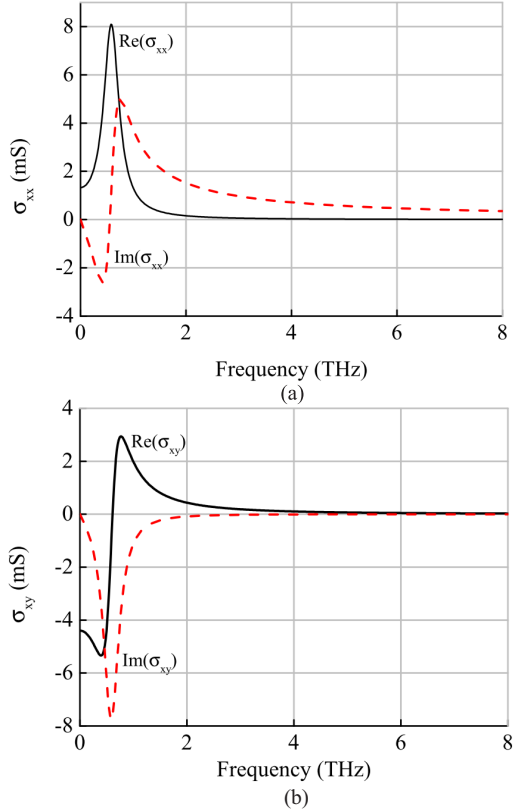}
\caption{Real and imaginary parts of components (a) $\sigma_{xx}$ and (b) 
$\sigma_{xy}$ of the graphene conductivity tensor, $\mu_c$ = 0.15 eV, $B_{0}$ = 0.61 T.}
\label{fig:Figcondu}
\end{figure}

Both the diagonal and the nondiagonal components of the graphene conductivity tensor depend on  field $B_{0}$. The chemical potential $\mu_c$ can be tuned by a bias voltage applied between the silicon substrate and the graphene with silica as a spacer. Frequency dependences of the real and imaginary parts of the tensor components $\sigma_{xx}$ and $\sigma_{xy}$  are shown in Fig. \ref{fig:Figcondu}.

In numerical calculus of SPP, graphene can be modelled as a bulk material with the conductivity tensor defined by $[\sigma_v]=[\sigma_s]/\Delta$ where $[\sigma_s]$ is the surface conductivity tensor \cite{Hanson}. Its components are given  by (\ref{eq:sig1}) and (\ref{eq:sig2}), and $\Delta=1$ nm is the thickness of the graphene. The artificial parameter $\Delta$ is used only for calculation purposes \cite{Transformation}.

For waveguides that are not magnetized,  graphene possesses  isotropic properties, so the conductivity model that we will use for numerical calculations are obtained from the tensor above, putting $B_0$ = 0 T. Thus,  $\sigma_{xy}$ = $\sigma_{yx}$ = 0 and $\sigma_{xx}$ = $\sigma_{yy}\neq0$, so one has the following expresion:
\begin{eqnarray}
\label{eq:sig3}
\sigma_{s} = \frac{2D}{\pi} \frac{i\omega - 1/\tau}{( \omega + i / \tau ) ^2},
\end{eqnarray}
\section{Graphene Resonator Analysis}
In the following, we shall use the same chemical potential for the waveguides $\mu_{cg}$ and for the resonator $\mu_{cr}$ ($\mu_{cg}$ = $\mu_{cr}$ = $\mu_{c}$. In the waveguides and in the graphene resonator,  TE-TM hybrid modes with all components of the electric and magnetic fields exist \cite {Luiz2015}. It can be shown, that the TM mode is the dominant one in the resonator \cite{wagner}.  
Therefore, approximately, one can define the radius $R$ of the resonator with dipole TM mode from the condition $2\pi R=\lambda_{spp}$ where $\lambda_{spp}$ is the wavelength of the TM SPP mode, therefore, $R=\lambda_{spp}/2\pi$. The phase constant $\beta_{spp}$ can be calculated by the dispersion relation for the TM SPP mode in an infinite graphene on a substrate \cite{Livro}:
\begin{equation}
\beta_{spp}=\frac{(1+\varepsilon_1)(\omega\hbar)^2}{2\alpha\mu_c\hbar c}\biggl(1-\frac {ga_{B}^2}{\omega^2}\biggr),
\label{eq:sig4}
\end{equation}
where $\alpha=e^2/(4\pi \varepsilon_0 \hbar c) \approx 0.007$ is the fine-structure constant, $\varepsilon_1$ is the dielectric constant of the substrate. From the relations $\beta_{spp}=2\pi/\lambda_{spp}$ and $R=\lambda_{spp}/2\pi$, one can find the radius of the resonator $R=1/\beta_{sp}$. Using $\beta_{spp}$ from (\ref{eq:sig4}), one comes to

\begin{equation}
\label{eq:sig5}
R\approx 8.3\times10^{40}{\frac{\mu_c}{(1+\varepsilon_1)(\omega_c^2-\omega_{B}^2)}},
\end{equation}
where $\omega_c$ is the central frequency of  frequency band of the device, $\omega_c$ and $\omega_{B}$ in THz, $R$ in nm. One can see that the radius of the resonator $R$ depends on the  frequencies $\omega_c$ and $\omega_{B}$, and also on $\mu_c$ and $\varepsilon_1$.

\begin{figure}[htbp]
\centering
\includegraphics[width=3in]{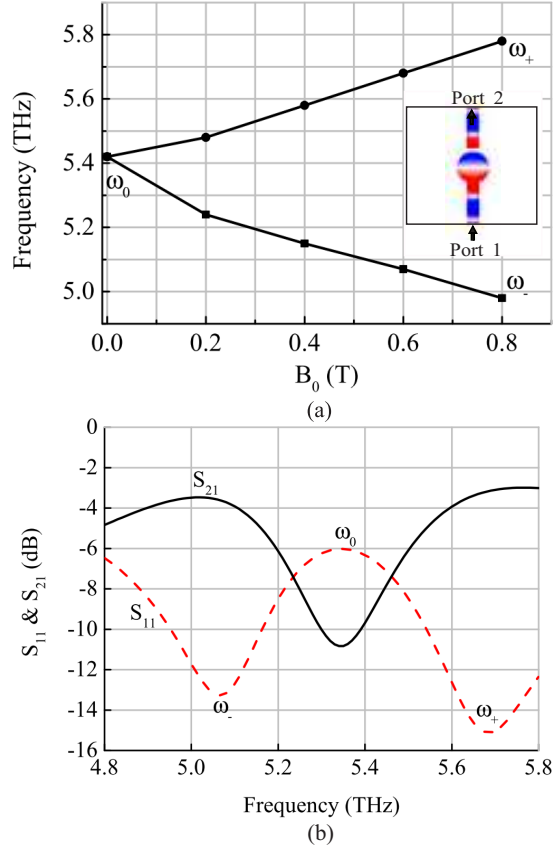}
\caption{(a) Frequencies of rotating modes versus DC magnetic field,  $R$ = 600 nm, $g$ = 5 nm, $w$ = 200 nm and $\mu_c$ = 0.15 eV; (b) Frequency responses of  resonator  showing resonances of rotating modes, $B_{0}$ = 0.61 T.}
\label{fig:desdo}
\end{figure}

The standing dipole mode in the nonmagnetized graphene resonator can be described as a sum of two counter-rotating modes with equal frequencies. When the resonator is magnetized, splitting of frequencies of the counter-rotating modes occurs \cite{wagner}. In order to analyze frequency dependence of resonances of the clockwise and anticlockwise rotating dipole modes on the magnetic field ${B}_0$, we calculated the structure with two waveguides coupled to the resonant cavity in Fig. \ref{fig:desdo}a. 
In Fig. \ref{fig:desdo}b, one can observe two peaks of the transmission coefficient and two  dips of reflection coefficient corresponding to the rotating modes with the frequencies $\omega_{+}$ and $\omega_{-}$. These calculus were performed for the feeding scheme with two waveguides and the gap of $g$ = $5$ nm between the guides and the resonator. The splitting of the modes $\omega_{+}$ and $\omega_{-}$ increases with enlargement of ${B}_0$.
\section{Operational Principle of Device}
To calculate the structure of the scattering matrix $S$ for the ON/OFF states of the device with two ports, we can use the commutation relation for the antiunitary element $\mathcal{T}\!\sigma$, that is, $R_{\sigma}S = S^TR_{\sigma}$ \cite{magnetic}, where $T$ indicates transposition and  $R_{\sigma}$ is the 2D representation of the operator $\sigma$. As a result, the matrix for the  device  can be written as follows:
\begin{equation}
S=
\left(\!\!
\begin{array}{cc}
S_{11}  & S_{12}\\
S_{21}  & S_{11} 
\end{array} \!\!\!
\right);
\label{eq:31}
\end{equation}
where $S_{22}=S_{11}$ due to symmetry. For the ideally matched two-port, one has $S_{22}=S_{11}=0$ and in the lossless case one comes to
\begin{equation}
S_{OFF}=
\left(\!\!
\begin{array}{cc}
1 & 0\\
0 & 1 
\end{array} \!\
\right);
\label{eq:32}
\end{equation}
for the OFF state,  and 
\begin{equation}
S_{ON}= 
\left(\!
\begin{array}{cc}
0 & 1 \\
e^{\,i\varphi\,} & 0     
\end{array} 
\right).
\label{eq:3}
\end{equation}
for the ON state. Notice that in the lossless matched case the four-port  can have only phase nonreciprocity and  matrix (\ref{eq:3}) describes the gyrator of Tellegen \cite{Tellegen}.

\begin{figure}[htbp]
\centering
\includegraphics[width=3.2in]{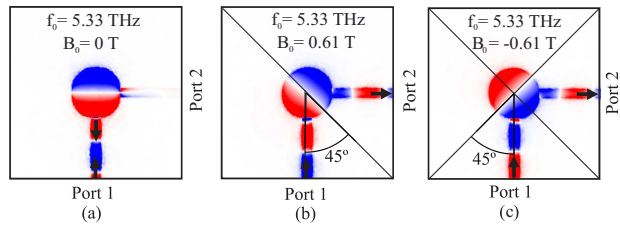}
\caption{$|E_z|$ field distribution in device,  (a) nonmagnetized (regime OFF). Magnetized (regime ON):  (b) and (c) transmission (1$\rightarrow$2), $B_{0}$ = $\pm$ 0.61 T, $\mu_c=0.15$ eV, $f_0$ = 5.33 THz.}
\label{fig:Figdipolo}
\end{figure}

In our theoretical model, we take into account only two lowest rotating dipole modes and the higher resonant modes are neglected. Without DC magnetic field $B_{0}$, the standing electromagnetic wave in the resonator is defined by the sum of the degenerate clockwise $\omega_{+}$ and anticlockwise $\omega_{-}$ rotating  modes.  The corresponding standing dipole mode does not excite the output port because it has the nodal plane in the center of the output waveguide and the eigenwave of this waveguide is symmetrical with respect to the plane of symmetry of the waveguide. Therefore, the input power is reflected back, Fig. \ref{fig:Figdipolo}a. This coresponds to regime OFF of the device. Thus, the dipole mode with the $90^\circ$  orientation of the output port allows one to provide ``a natural'' OFF state without graphene  magnetization.

For the magnetized resonator, the degeneracy of the clockwise $\omega_{+}$ and anticlockwise $\omega_{-}$ rotating dipole modes is removed. In this case, the sum of these modes at a frequency $\omega_{c}\approx (\omega_{+} + \omega_{-})/2$ produces also a standing wave but this dipole is oriented at the angle $\phi$ with respect  to the input port. By choosing the parameters of the resonator, one can achieve $\phi$ = $\pm$ 45$^{\circ}$ so that in port 1 and port 2 the fields will have equal amplitudes. If the impedances of the waveguides and the resonator are matched, there will be a complete transmission from the input port to the output port (i.e. regime ON) as shown in Fig. \ref{fig:Figdipolo}b,c.
\section{Parametric analysis}
The characteristics of the our device depend on several principal parameters, such as diameter of the resonator, resonator-waveguide coupling, biasing DC magnetic field, chemical potential. The main  results of our analysis concerning the influence of different parameters of the structure on its characteristics are summarized below.
\subsection{Radius of Resonator}
\begin{figure}[htbp]
\centering
\includegraphics[width=3.4in]{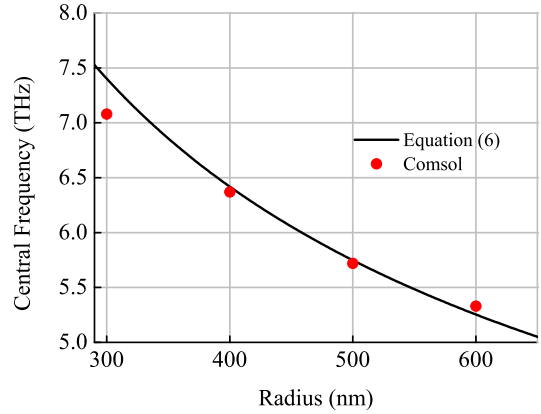}
\caption{Central frequency of device versus radius $R$, $g$ = 5 nm, $w$ = 300 nm, $\mu_c$ = 0.15 eV and $B_0$ = 0.61 T.}
\label{fig:freq_raio}
\end{figure}

\begin{figure}[htbp]
\centering
\includegraphics[width=3.4in]{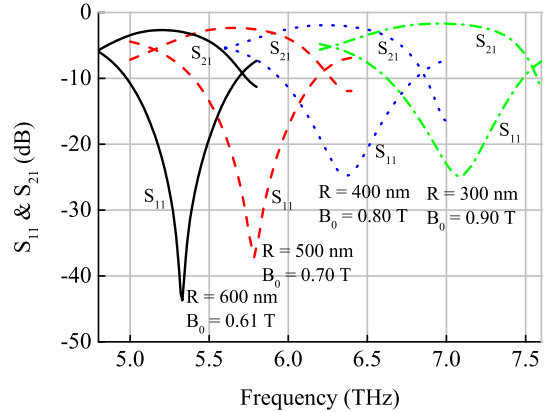}
\caption{Frequency response of the device with different $R$,  $g$ = 5 nm, $w$ = 300 nm, $\mu_c$ = 0.15 eV .}
\label{fig:Raio}
\end{figure}

\begin{figure}[htbp]
\centering
\includegraphics[width=3.4in]{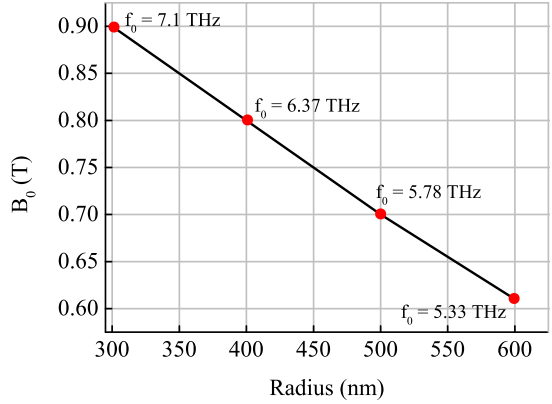}
\caption{Optimal magnetic field of device versus radius $R$, $g$ = 5 nm, $w$ = 300 nm, $\mu_c$ = 0.15 eV.}
\label{fig:raio_campo}
\end{figure}

The radius $R$ of the graphene resonator calculated by (\ref{eq:sig5}) and obtained from Comsol simulations are in a good agreement (see Fig. \ref{fig:freq_raio}). To analyze its influence on the performance of the switch, we varied $R$ between 350 and 600 nm, keeping  the parameters $w$ = 300 nm, $g$ = 5 nm and $\mu_c$ = 0.15 eV  fixed. The switch frequency responses for different $R$ and optimal magnetic field are plotted in Fig.\ref{fig:Raio}.  The reflection levels are better than -20 dB and the insertion losses are about $-(2\div3)$dB. One can see in Fig. \ref{fig:raio_campo} that with increase of the radius $R$,  the DC magnetic field decreases almost linearly. 

\subsection{Waveguide Width}
\begin{figure}[htbp]
\centering
\includegraphics[width=3.4in]{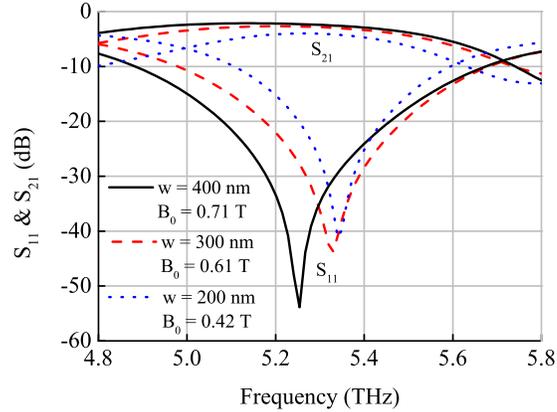}
\caption{Frequency response of the device with $g$ = 5 nm, $\mu_c$ = 0.15 eV, $R$ = 600 nm for different $w$.}
\label{fig:Largura}
\end{figure}

\begin{figure}[htbp]
\centering
\includegraphics[width=3.4in]{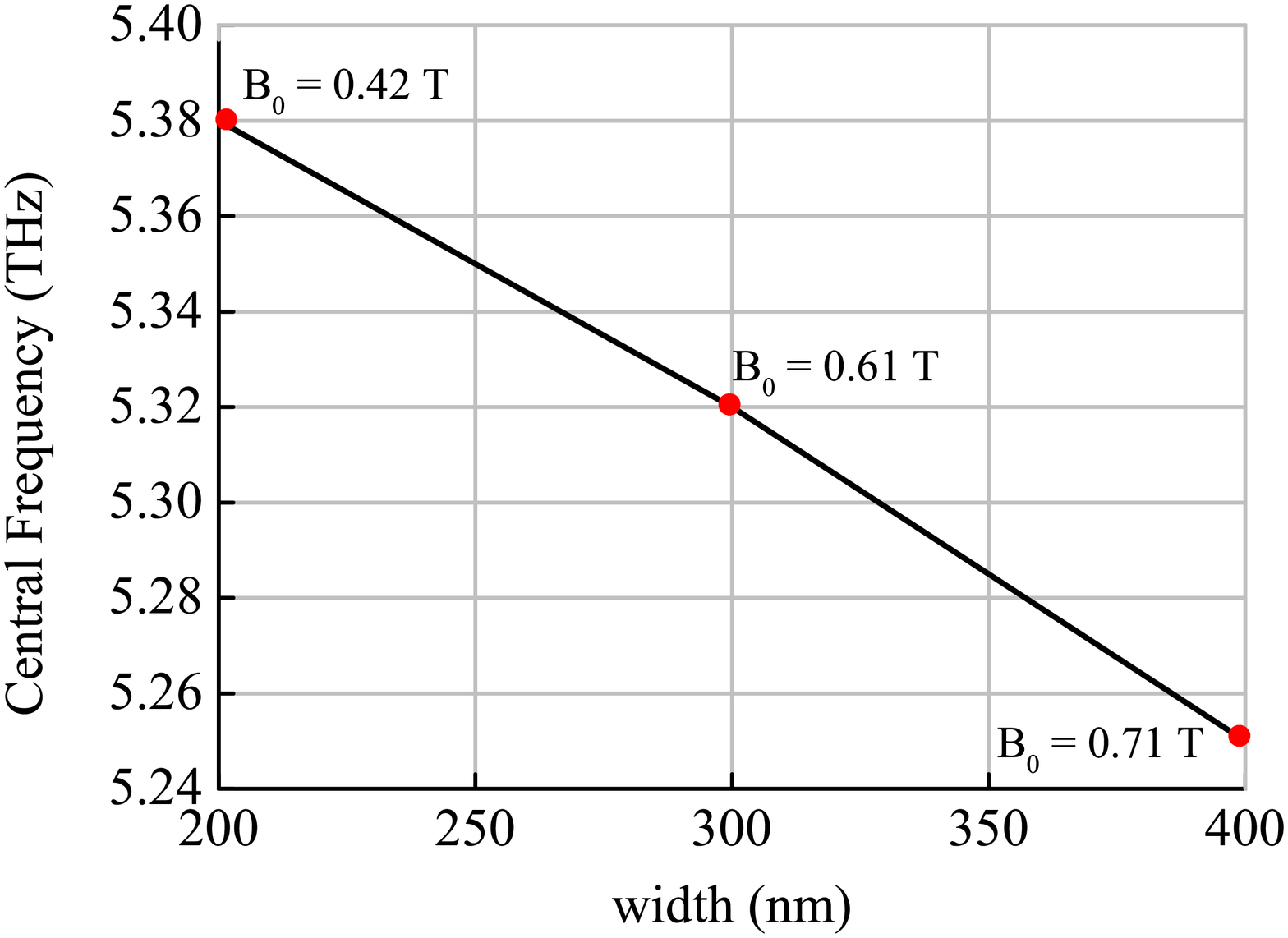}
\caption{Central frequency of device for different widths $w$, $g$ = 5 nm, $R$ = 600 nm, $\mu_c$ = 0.15 eV.}
\label{fig:freq_largura}
\end{figure}

\begin{figure}[htbp]
\centering
\includegraphics[width=3.4in]{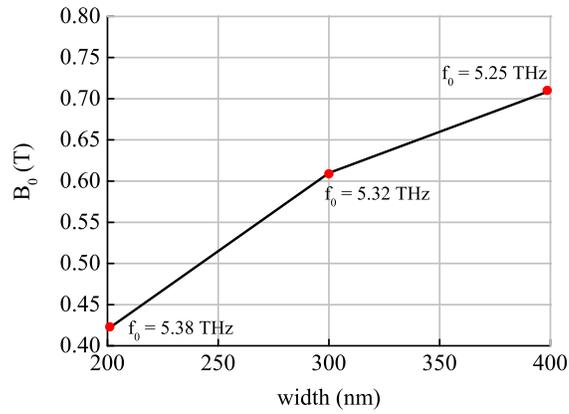}
\caption{Optimal magnetic field of device versus $w$, $R$ = 600 nm, $g$ = 5 nm, $\mu_c$ = 0.15 eV.}
\label{fig:campo_largura}
\end{figure}

The width of the waveguides $w$ defines the strength of  coupling of the resonator and waveguides.
The device frequency responses for the 200, 300 and 400 nm widths, keeping all other  parameters fixed with the values $R$ = 600 nm, $g$ = 5 nm and $\mu_c$ = 0.15 eV, are shown in Fig. \ref{fig:Largura}. The resonance frequency with increase of the width of the graphene waveguides shifts slightly to smaller values. In Fig. \ref{fig:freq_largura} and in 
Fig. \ref{fig:campo_largura} we see the variation of the resonant frequency and the optimized magnetic field with the width. For the 400 nm width, one can observe  better characteristics but at the expense of higher magnetic field (see Fig. \ref{fig:Largura}).

\subsection{Waveguide-Resonator Gap}
\begin{figure}[htbp]
\centering
\includegraphics[width=3.4in]{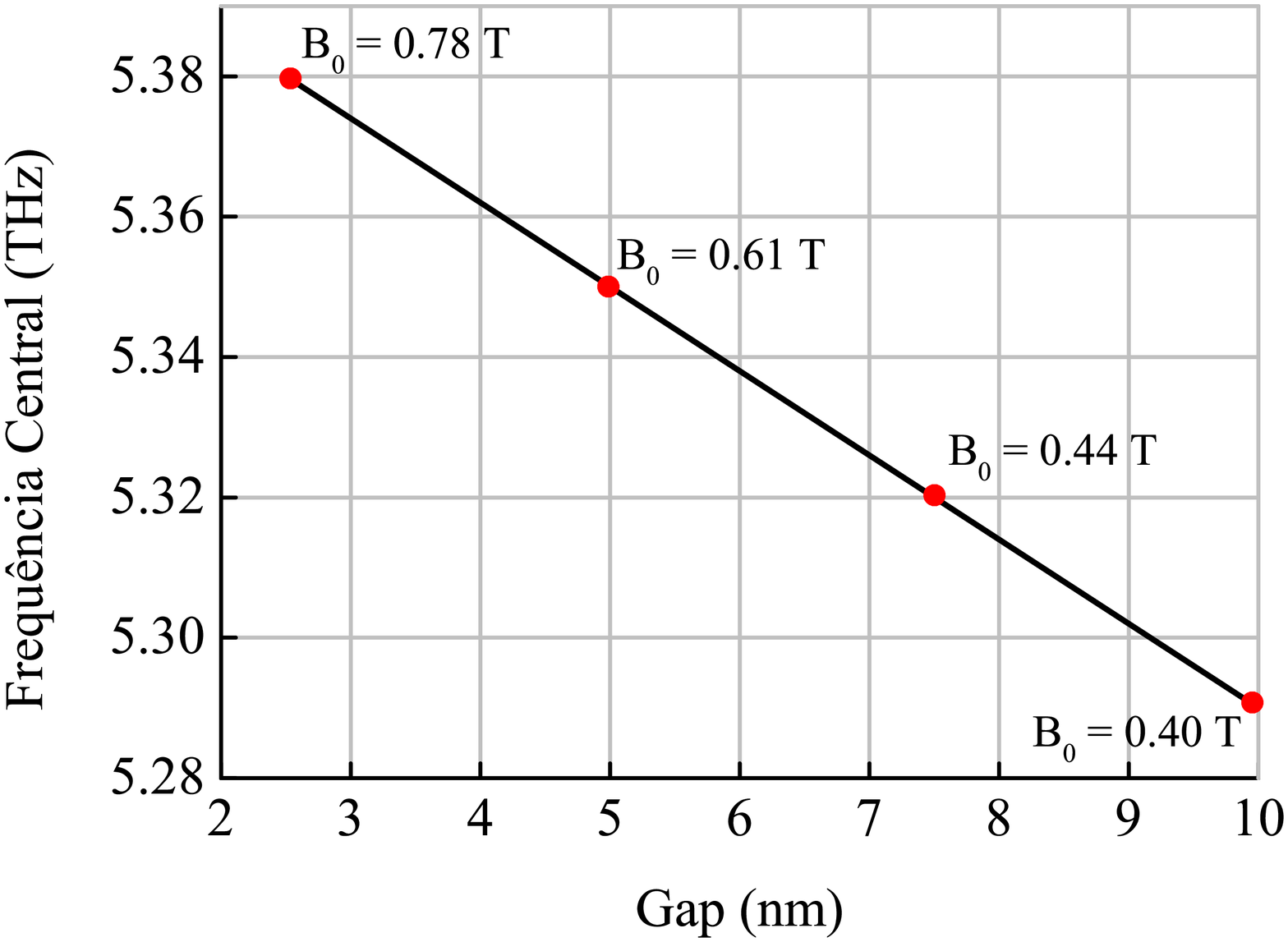}
\caption{Central frequency of device for different values of gap $g$, $w$ = 300 nm, $R$ = 600 nm, $\mu_c$ = 0.15 eV.}
\label{fig:gap_freq}
\end{figure}

\begin{figure}[htbp]
\centering
\includegraphics[width=3.4in]{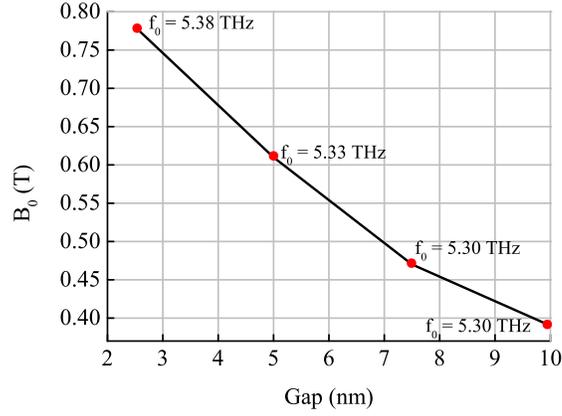}
\caption{Optimal magnetic field of device for different values of gap $g$, $w$ = 300 nm, $R$ = 600 nm, $\mu_c$ = 0.15 eV.}
\label{fig:gap_B}
\end{figure}

\begin{figure}[htbp]
\centering
\includegraphics[width=3.4in]{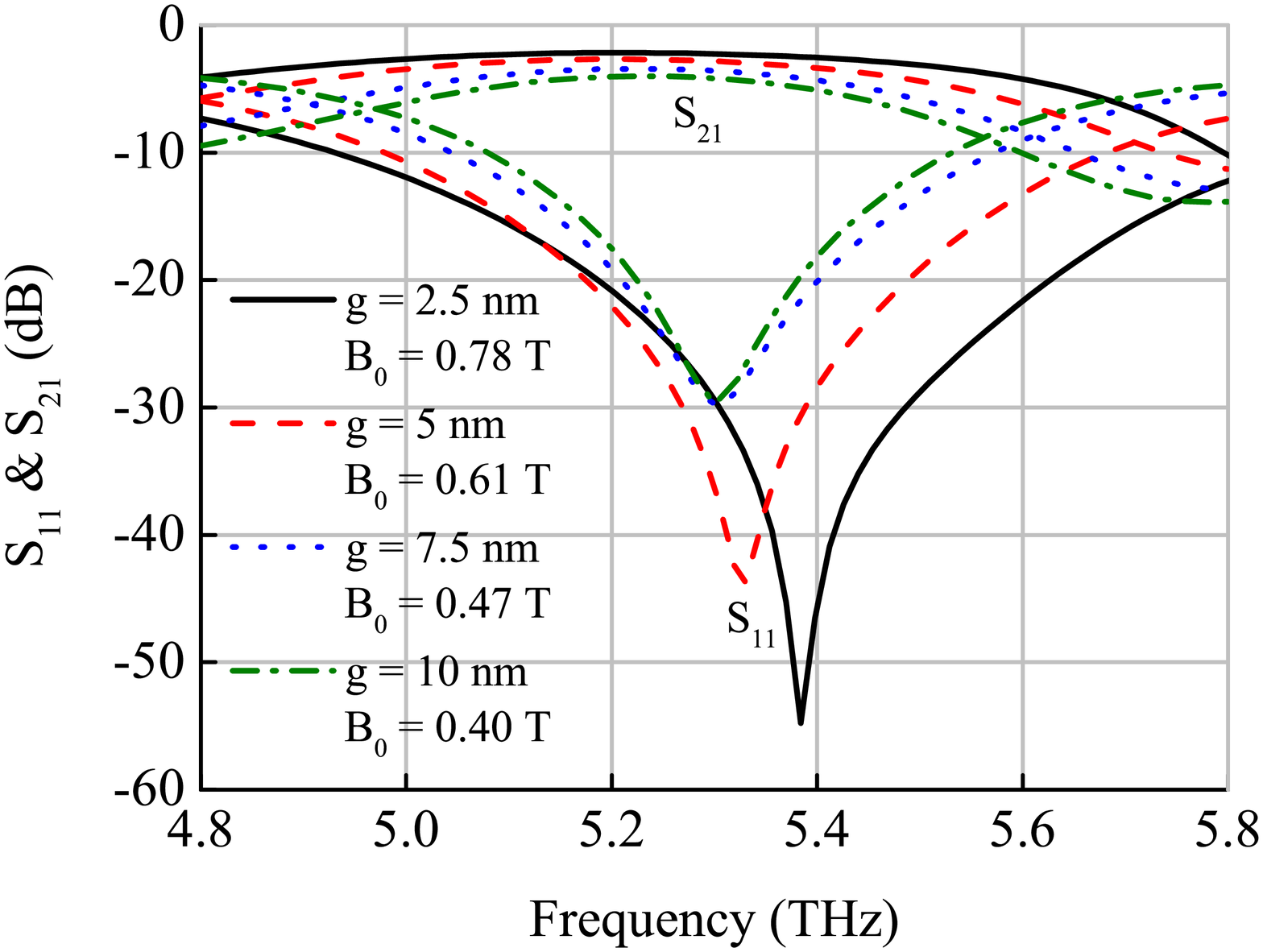}
\caption{Frequency response of the device with $w$ = 300 nm, $\mu_c$ = 0.15 eV for different $g$.}
\label{fig:gap}
\end{figure}

The gap $g$ also defines coupling of the resonator and waveguides. In our analysis, the gap was varied from 2.5 nm to 10 nm with the fixed  $R$ = 600 nm, $w$ = 300 nm and $\mu_c$ = 0.15 eV. From Fig. \ref{fig:gap_freq} and 
Fig. \ref{fig:gap_B} one can see that the gap has influence on the frequency and DC magnetic field. The frequency responses of the device along with the optimal magnetic fields are shown in  Fig. \ref{fig:gap}.  

\subsection{Chemical Potential}
\begin{figure}[htbp]
\centering
\includegraphics[width=3.4in]{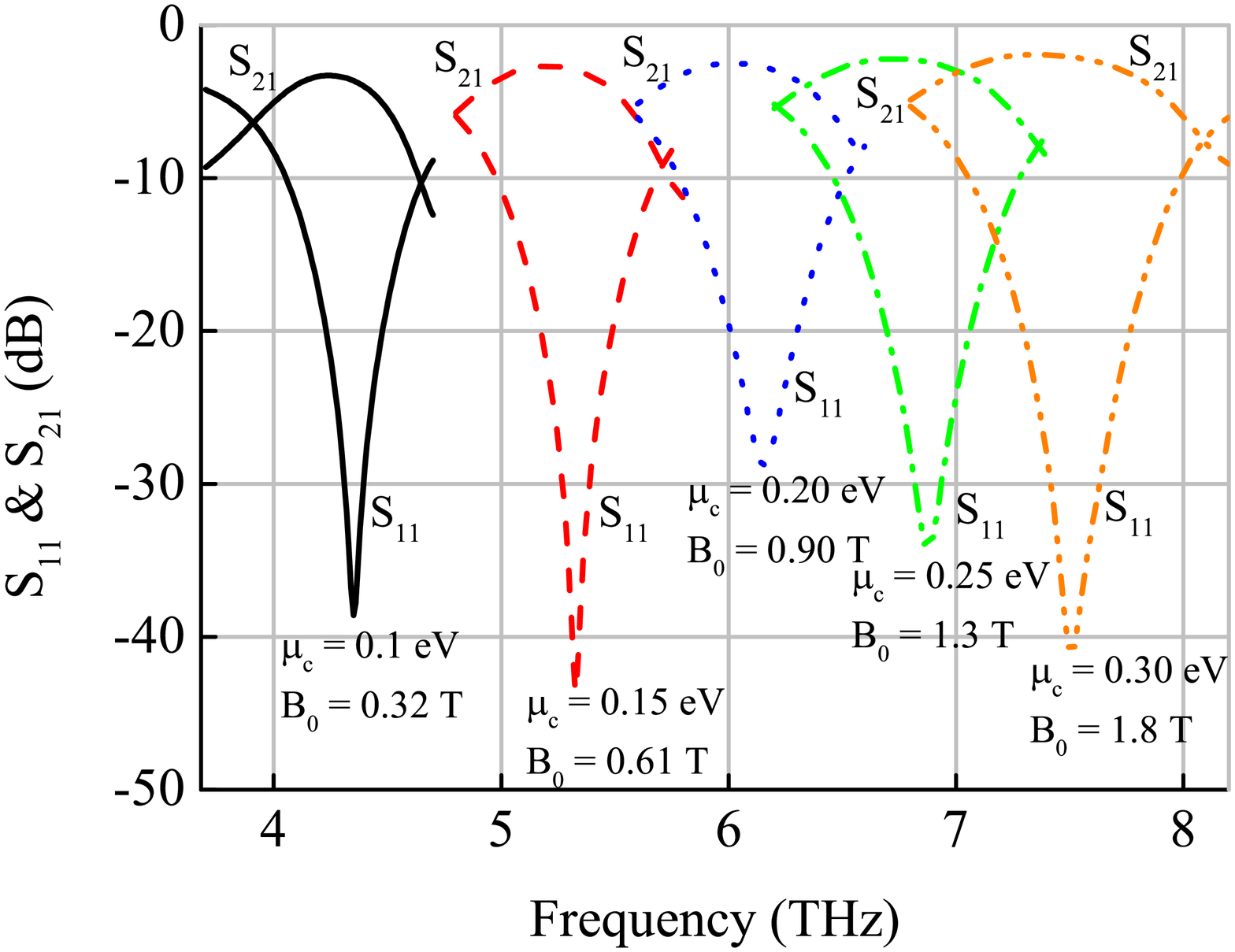}
\caption{Frequency response of the device with $g$ = 5 nm, $R$ = 600 nm, $w$ = 300 nm for different chemical potential.}
\label{fig:Energy}
\end{figure}

\begin{figure}[htbp]
\centering
\includegraphics[width=3.4in]{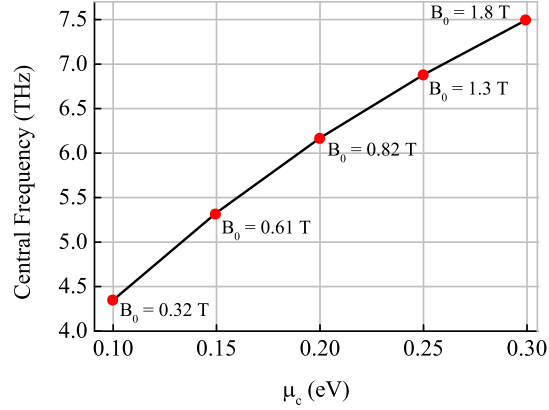}
\caption{Optimal magnetic field $B_0$ for device with $g$ = 5 nm, $R$ = 600 nm, $w$ = 300 nm for different chemical potential.}
\label{fig:freq_Fermi}
\end{figure}

Varying the chemical potential of graphene produced by an external electric field changes the density of the charge carriers of the material. In this way, it is possible to dynamically control the  resonant frequency of the device. Influence of the chemical potential on the resonance frequencies for the structure with $R$ = 600 nm, $g$ = 5 nm, $w$ = 300  nm   can be analyzed in 
Fig. \ref{fig:Energy}. Increase of the optimal magnetic field with increasing the chemical potential in 
Fig. \ref{fig:freq_Fermi} can be explained by the relation  for the cyclotron frequency $\omega_B = {eB_{0}v_{F}^{2}}/{\mu_c}$.  
\subsection{Effect of Sign Changing Field}
Switching  direction of the external DC magnetic field from  $+{B}_0$ to $-{B}_0$ leads to transposition of the scattering matrix, i.e.  to a  change in  characteristics of the device. In order to evaluate this change, the frequency responses of the device have been calculated for a certain direction of the external DC magnetic field $+{B}_0$. If this field is switched to the opposite one, that is $+{B}_0 \rightarrow -{B}_0$, the rotation modes with $\omega_{+}$ and $\omega_{-}$ are interchanged,   Fig. \ref{fig:desdo}a.  Thus, the standing dipole mode in the resonator for $+{B}_0$ corresponds to the odd symmetry of the electric field with respect to the geometrical plane of symmetry $\sigma$ 
(Fig. \ref{fig:Figdipolo}b), but for $-{B}_0$, this mode is defined by the even symmetry (Fig. \ref{fig:Figdipolo}c). It means, that the wave passing through the device with $-{B}_0$ will have a different phase shift in comparison with  case of $+{B}_0$ (see (\ref{eq:3})) and slightly different characteristics (see (\ref{eq:31})).
%
%
\begin{figure}[htbp]
\centering
\includegraphics[width=3.4in]{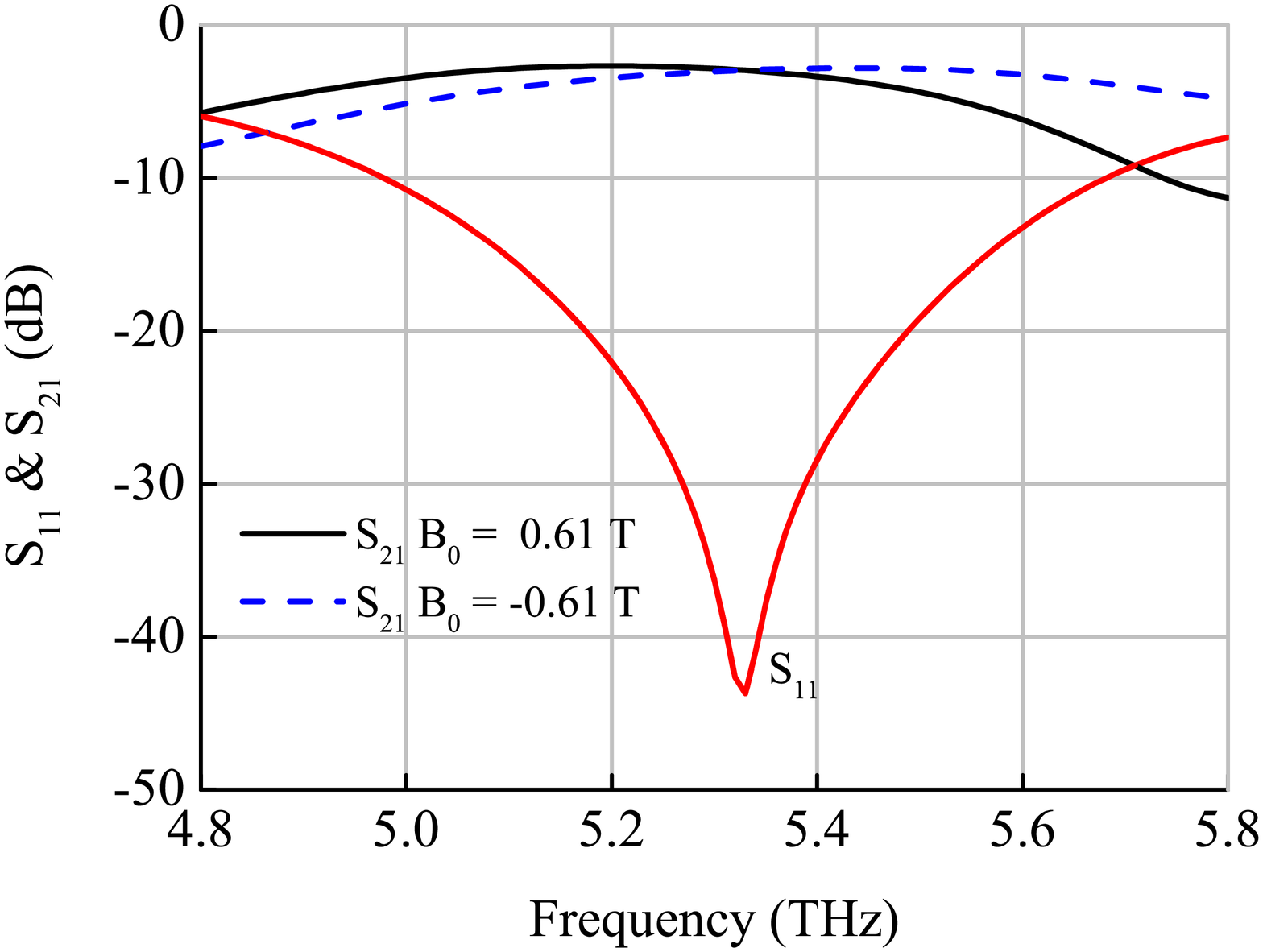}
\caption{Effect on the device frequency response  due to change in orientation of $B_0$ with $\mu_c$ = 0.15 eV, $w$ = 300 nm, $g$ = 5 nm and $R$ = 600 nm.}
\label{fig:B}
\end{figure}
\section{Switch Design}
Below we present an example of the switch-filter project.
\subsection{Switching by DC magnetic Field}
\begin{figure}[htbp]
\centering
\includegraphics[width=3.4in]{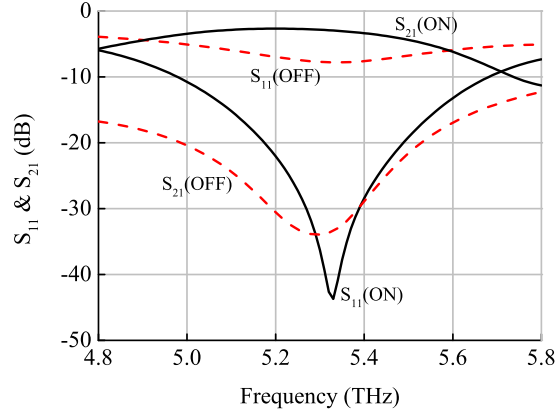}
\caption{Frequency response for the device in the ON and OFF states for ON ($B_0$ = 0.61 T) and OFF ($B_0$ = 0 T), $g$ = 5 nm, $w$ = 300 nm and $\mu_c$ = 0.15 eV.}
\label{fig:ONOFF61}
\end{figure}

Fig. \ref{fig:ONOFF61} shows the frequency response of the device with and without magnetization, i.e for the ON and OFF states. We can see that the application of the DC magnetic field $B_0$ = 0.61 T produces the rotation of the standing dipole by $45^\circ$ causing the wave to follow from port 1 to port 2. For the ON state the reflection level at the central frequency is around -40 dB and insertion loss is  around -2 dB, with  Q-factor $\approx$ 7.8 and the half power bandwidth $HPBW$ = 12.7$\%$. If the magnetic field  is switched OFF, the dipole will return to its natural state and  port 2 will be blocked. In the OFF state,  the reflection coefficient  is about -7 dB (the relatively low value of the reflection means that a part of the electromagnetic energy is absorbed in the resonator) and the insertion loss is around -33 dB. It the ON/OFF ratio is better than 20 dB in the frequency band of the device.
\subsection{Switching by Chemical Potential} 
\begin{figure}[htbp]
\centering
\includegraphics[width=3.3in]{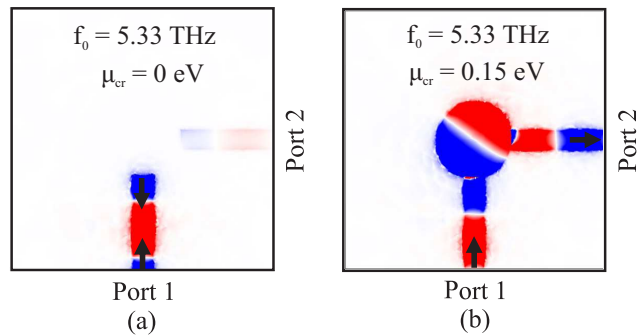}
\caption{$|E_z|$ field for the device in the ON/OFF states; (a) OFF ($B_0$ = 0.61 T, $\mu_{cg}$ = 0.15 eV and $\mu_{cr}$ = 0 eV); (b) for ON ($B_0$ = 0.61 T, $\mu_{cg}$ = $\mu_{cr}$ = 0.15 eV) and , $g$ = 5 nm, $w$ = 300 nm.}
\label{fig:61mi00}
\end{figure}

\begin{figure}[htbp]
\centering
\includegraphics[width=3.4in]{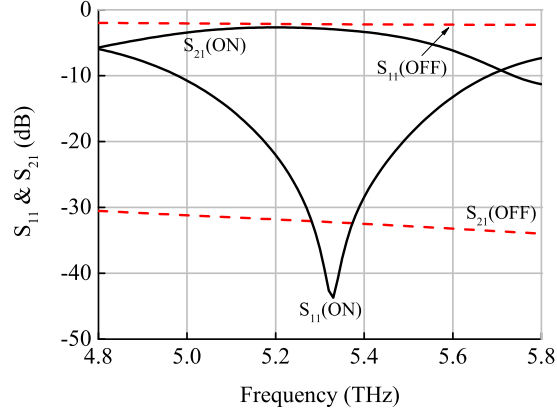}
\caption{Frequency response for the device in the ON/OFF states for ON ($B_0$ = 0.61 T, $\mu_{cg}$ = $\mu_{cr}$ = 0.15 eV) and OFF ($B_0$ = 0.61 T, $\mu_{cg}$ = 0.15 eV and $\mu_{cr}$ = 0 eV), $g$ = 5 nm, $w$ = 300 nm.}
\label{fig:ONOFF0061}
\end{figure}

For graphene  biasing  by electric field, a special polisilicon layer  parallel to graphene  can be used as a gate electrode  to provide DC voltage between this electrode and graphene \cite{Gomez}. 
Fig. \ref{fig:61mi00} shows $|E_z|$ field distribution for the ON/OFF states, considering fixed $B_0$ = 0.61 T and the chemical potential of guide $\mu_{cg}$ = 0.15 eV, switching the chemical potential in the resonator $\mu_{cr}$ from 0 eV to 0.15 eV. The frequency response for the device is shown in  Fig. \ref{fig:ONOFF0061}, where the reflection coefficient is about -2 dB (it means that almost all  energy is reflected) and the  the insertion loss is higher than -30 dB for the OFF state. For the ON state we have insertion losses around -2 dB with reflection level of -40 dB. 

\subsection{Tuning  by Chemical Potential} 
\begin{figure}[htbp]
\centering
\includegraphics[width=3.3in]{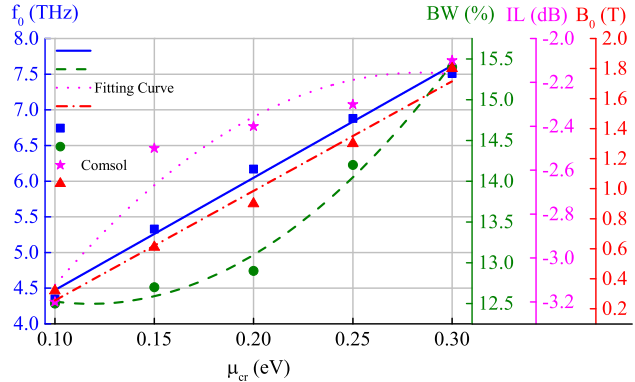}
\caption{Influence of chemical potential on central frequency (circles), bandwidth (asterisk), insertion losses (x) and optimal magnetic field $B_0$ (triangle).}
\label{fig:yyyy}
\end{figure}

\begin{figure}[htbp]
\centering
\includegraphics[width=3.4in]{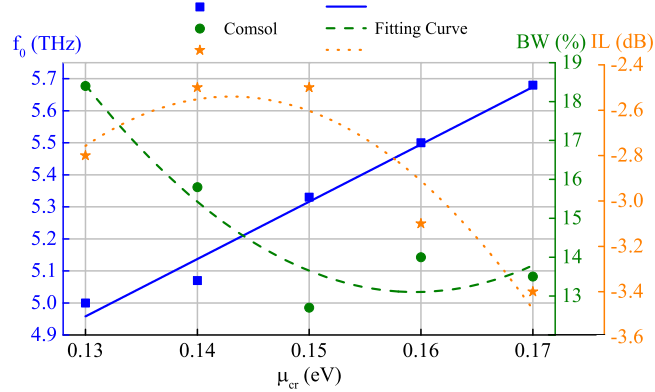}
\caption{Influence of chemical potential on central frequency (square), bandwidth (circles) and insertion loss (star),  $B_0$ = 0.61 T is fixed.}
\label{fig:yyy}
\end{figure}

Fig. \ref{fig:yyyy} shows  influence of the chemical potential variation from 0.1 to 0.3 eV on the central frequency ($f_0$), bandwidth (BW) and insertion loss (IL) of the device. In our analysis, the external magnetic field $B_0$ was optimized for each chemical potential value. With  increase of the chemical potential, the device operating frequency shifts from 4.5 THz to 7.5 THz,  the bandwidth is enlarged and the insertion loss is reduced. However, it is achieved at the expense of the higher biasing magnetic field.
We also analysed  influence of  the chemical potential variation from 0.13 eV to 0.17 eV on the device parameters, maintaining the  magnetic field $B_0$ = 0.61 T fixed. The frequency tuning in this case  is possible  from 5 THz to 5.7 THz (see Fig. \ref{fig:yyy}) with  a small degradation of  the device bandwidth and the  insertion loss. Thus,  with a fixed DC magnetic field, a fine tuning of the central frequency by chemical potential is possible.
\section{Conclusion}
In our work we propose and analyze a graphene-based  device acting as a switch and filter.  The ON state of the device corresponds to the magnetized graphene.  The OFF state can be achieved in  two different ways. One of them is switching the external magnetic field to zero, and the second way is switching the electric field (i.e. the chemical potential) to zero. We demonstrated also a possibility of the  dynamic adjustment of the device  in a wide frequency range.  The suggested structure  with  high ON/OFF ratio, filtering properties and a high tunability and has a potential of application for  THz and IR switches and modulators. The presented results can serve as guidelines for the design of such devices.
\section*{Acknowledgment}
The authors would like to thank the Brazilian agency National Counsel of Technological and Scientific Development for the  financial support.
%



@inproceedings{kour2014real,
  title={Real-time segmentation of on-line handwritten arabic script},
  author={Kour, George and Saabne, Raid},
  booktitle={Frontiers in Handwriting Recognition (ICFHR), 2014 14th International Conference on},
  pages={417--422},
  year={2014},
  organization={IEEE}
}

@inproceedings{kour2014fast,
  title={Fast classification of handwritten on-line Arabic characters},
  author={Kour, George and Saabne, Raid},
  booktitle={Soft Computing and Pattern Recognition (SoCPaR), 2014 6th International Conference of},
  pages={312--318},
  year={2014},
  organization={IEEE}
}

@article{hadash2018estimate,
  title={Estimate and Replace: A Novel Approach to Integrating Deep Neural Networks with Existing Applications},
  author={Hadash, Guy and Kermany, Einat and Carmeli, Boaz and Lavi, Ofer and Kour, George and Jacovi, Alon},
  journal={arXiv preprint arXiv:1804.09028},
  year={2018}
}

\begin{thebibliography}{1}
		\bibitem{graphene} K. S. Novoselov, and A. K. Geim, ``Electric field effect in atomically thin carbon films'', {\it Science, American Association for the advancement of Science}, vol. 306, no. 5996, pp. 666-669, Out. 2004.
		\bibitem{R1}
		R. Wang, X. G. Ren, Z. Yan, L. J. Jiang, W. E. I. Sha and G. C. Shan, ``Graphene based functional devices: a short review'', {\it Front. Phys.}, 
vol. 14, no. 1, pp. 13603-13623, Oct. 2019.
       \bibitem{R2}
        S. Huang, C. Song, G. Zhang and H. Yan, ``Graphene plasmonics: physics and potential applications'', {\it Nanophotonics}, vol. 6, pp. 1191–1204, Oct. 2016.
       \bibitem{R3}
       P. Avouris and M. Freitag, ``Graphene Photonics, Plasmonics, and Optoelectronics'', {\it IEEE Journal of Selected Topics in Quantum Electronics}, vol. 20, no. 1, pp. 72-83, Feb. 2014.
		\bibitem{Switches1} J. Foster, G. Edmiston, M. Thomas and A. Neuber, ``High power microwave switching utilizing a waveguide spark gap'', {\it Review of Scientific Instruments}, vol. 79, no. 11, pp. 114701--114705, Nov. 2008.        
        
        \bibitem{Switches2} R. L. Espinola,  M. C. Tsai, J. T. Yardley and R. M. Osgood, Jr., ``Fast and low-power thermooptic switch on thin silicon-on-insulator'',  {\it IEEE Photonics Technology Letters}, vol. 15, no. 10, pp. 1366--1368, Sep. 2003.
        \bibitem{S3} Yu, Z., Sun, X, ``Acousto-optic modulation of photonic bound state in the continuum'', {\it Light Sci. Appl.}, vol. 9, no. 1, pp. 1--9, Jan. 2020.
        \bibitem{S4} Abu Naim R. Ahmed, Sean Nelan, Shouyuan Shi, Peng Yao, Andrew Mercante and Dennis W. Prather, ``Subvolt electro-optical modulator on thin-film lithium niobate and silicon nitride hybrid platform'', {\it Opt. Lett.}, 
vol. 45, no. 5, pp. 1112-1115, Mar. 2020.
        \bibitem{S5} E. A. Tsygankov, S. A. Zibrov, A. S. Zibrov, M. I. Vaskovskaya, D. S. Chuchelov, V. V. Vassiliev,  V. L. Velichansky, S. V. Petropavlovsky, V. P.  Yakovlev, ``Single magneto-optical resonance in a modulated RF field'', {\it Phys. Rev. A}, 
vol. 99, no. 6, pp. 063835-063842, Jun. 2019.
        \bibitem{Mach-Zehnder} R.G. Heideman, P.V. Lambeck, ``Remote opto-chemical sensing with extreme sensitivity: design, fabrication and performance of a pigtailed integrated optical phase-modulated Mach–Zehnder interferometer system'', {\it Sensors and Actuators B}, vol. 61, no. 3, pp. 100--127, Dec. 1999.
		\bibitem{photonic-crystal} V. Dmitriev, G. Portela and D. Zimmer, ''Possible mechanisms of switching in symmetrical two-ports based on 2D photonic crystals with magneto-optical resonators'', {\it Optics Letters}, 
vol. 38, no. 20, pp. 4040--4043, 
Oct. 2013.
        \bibitem{Feng} F.Cheng, ``A tunable high-efficiency optical switch based on graphene coupled photonic crystals structure'', {\it Journal of Modern optics}, vol. 64, no. 15, pp. 1531-1537, Mar. 2017.
        \bibitem{Zhang} Z. Zhang, J. Yang, X. He, Y. Han, J. Zhang, J. Huang, D. Chen, S. Xu, ``All-optical multi-channel switching at telecommunication wavelengths based on tunable plasmon-induced transparency'', {\it Optics Communications}, vol. 425, no. 6, pp. 196--203, Oct. 2018.
        \bibitem{Luo} L. Luo, K. Wang, C. Ge, K. Guo, F. Shen, Z. Yin, Z. Guo, ``Actively controllable terahertz switches with graphene-based nongroove gratings'', {\it Photon. Res.}, vol. 5, no. 6, pp. 604--611, Dec. 2017.
	    \bibitem{Ali} A. Farmani, A. Zarifkar, M. H. Sheikhi, M. Miri, ``Design of a tunable graphene plasmonic-on-white graphene switch at infrared range'', {\it Superlattices and Microstructures},
vol. 112, no. 1, pp. 404--414, Dec. 2017.
        \bibitem{Sun} S. Jian-Zhong, Z. Le, G. Fei, ``Switching terahertz waves with graphene-integrated split-ring resonator'', {\it Optik}, vol. 127, no. 19, pp. 8096--8102, 
May. 2016.
        \bibitem{Guo} G. Zhongyi et al. ``Actively tunable terahertz switches based on subwavelength graphene waveguide'', {\it Nanomaterials}, 
vol. 8, no. 9, pp. 665--676, Aug. 2018.
	    \bibitem{y} A. Farmani, M. Yavarian, A. Alighanbari, M. Miri and M. H. Sheikhi, ``Tunable graphene plasmonic Y-branch switch in the terahertz region using hexagonal boron nitride with electric and magnetic biasing'', {\it Appl. Opt.}, vol. 56, no. 32, pp. 8931--8940, Nov. 2017.
	    \bibitem{MP} M. Heidari and V. Ahmadi, ``Design and Analysis of a Graphene Magneto-Plasmon Waveguide for Plasmonic Mode Switch'', {\it IEEE Access}, vol. 7, no. 1, pp. 43406--43413, Apr. 2019.
	    \bibitem{Fi1} Hamed Arianfard, Bahareh Khajeheian, Rahim Ghayour, ``Tunable band (pass and stop) filters based on plasmonic structures using Kerr-type nonlinear rectangular nanocavity'', {\it Optical Engineering}, 
        vol. 56, no. 12, pp. 121902--121907, Dec. 2017.
        \bibitem{Fi2} Z. Z. Jiang and L. J. Sheng, ``Terahertz Band-stop Filter based on graphene cavity'', {\it Micro and Nano Letters}, vol. 13, no. 3, pp. 374--377, Mar. 2018.
	    \bibitem{comsolSite} http://www.comsol.com.br
        \bibitem{magnetic} A. Barybin and V. Dmitriev, ``Modern Electrodynamics and Coupled Mode Theory: Application to Guided-Wave Optics'', {\it Princeton}, NJ, USA: Rinton Press, 2002.
	   \bibitem{sigma} L. Giampiero, H. W. George,  A. Rodolfo and  B. Paolo,  ``Semiclassical spatially dispersive intraband conductivity tensor and quantum capacitance of graphene'', {\it Phys. Rev. B}, vol. 87, no. 11, pp. 115429-115440, Mar. 2013.
	   \bibitem{inter} Y. V. Bludov, A. Ferreira, N. M. R. Peres and M. I. Vasilevskiy, ``A Primer on Surface Plasmon-Polarintons in graphene'', 
 {\it Int. Journal of Mod. Phys. B}, 
vol. 27, no. 10, pp. 1341001--1341075, Apr. 2013.
      \bibitem{tau} A. Principi and G. Vignale, ``Intrinsic lifetime of Dirac plasmons in graphene'', {\it Phys. Rev. B}, vol. 88, no. 19, pp. 195405--195420, Nov. 2013.
        \bibitem{Hanson} G. W. Hanson, ``Dyadic Green's functions and guided surface waves for a surface conductivity model of graphene'', {\it Journal of Appl. Phys.}, vol. 103, no. 6, pp. 064302-064400, Jan. 2008.
       \bibitem{Transformation} A. Vakil and N. Engheta, ``Transformation optics using graphene'', {\it Science}, vol. 332, no. 6035, pp. 1291--1294, 2011.
        \bibitem{Luiz2015} L. G. C. Melo, ``Theory of magnetically controlled low-terahertz surface plasmon-polariton modes in graphene-dielectric structures'', {\it Journal of the Opt. Socievty of America B}, vol. 32, no. 12, pp. 2467--2477, Dec. 2015.
        \bibitem{wagner} V. Dmitriev and W. Castro, ``Dynamically Controllable Terahertz Graphene Y-Circulator'', {\it IEEE Transactions on Magnetics}, vol. 55, no. 2, pp. 1--12, Feb. 2019.
         \bibitem{Livro} P. A. D. Gonçalves, and N. M. R. Peres, ``An introduction to graphene plasmonics'', {\it World Scientific}, New Jersey, 2016.
        \bibitem{Tellegen} B. D. H. Tellegen, ``The gyrator, a new electric network element'', {\it Philips Res. Rep.}, vol. 3, no. 2, pp. 81--101, Abr. 1948.
       \bibitem{Gomez} Gomez-Diaz, J., Moldovan, C., Capdevila, S. et al., ``Self-biased reconfigurable graphene stacks for terahertz plasmonics'', {\it Nat. Commun.}, vol. 6, no. 6334, pp. 1--8, Mar. 2015.

\bibitem{kour2014real}
George Kour and Raid Saabne.
\newblock Real-time segmentation of on-line handwritten arabic script.
\newblock In {\em Frontiers in Handwriting Recognition (ICFHR), 2014 14th
  International Conference on}, pages 417--422. IEEE, 2014.

\bibitem{kour2014fast}
George Kour and Raid Saabne.
\newblock Fast classification of handwritten on-line arabic characters.
\newblock In {\em Soft Computing and Pattern Recognition (SoCPaR), 2014 6th
  International Conference of}, pages 312--318. IEEE, 2014.

\bibitem{hadash2018estimate}
Guy Hadash, Einat Kermany, Boaz Carmeli, Ofer Lavi, George Kour, and Alon
  Jacovi.
\newblock Estimate and replace: A novel approach to integrating deep neural
  networks with existing applications.
\newblock {\em arXiv preprint arXiv:1804.09028}, 2018.

\end{thebibliography}

\end{document}